\input harvmac
\newcount\yearltd\yearltd=\year\advance\yearltd by 0

\let\includefigures=\iftrue
\input epsf

\newcount\figno
\figno=0
\def\fig#1#2#3{
\par\begingroup\parindent=0pt\leftskip=1cm\rightskip=1cm\parindent=0pt
\baselineskip=11pt
\global\advance\figno by 1
\midinsert
\epsfxsize=#3
\centerline{\epsfbox{#2}}
\vskip 12pt
{\bf Figure \the\figno:} #1\par
\endinsert\endgroup\par
}
\def\figlabel#1{\xdef#1{\the\figno}}


\def\journal#1&#2(#3){\unskip, \sl #1\ \bf #2 \rm(19#3) }
\def\andjournal#1&#2(#3){\sl #1~\bf #2 \rm (19#3) }

\def\frac#1#2{{#1\over#2}}

\def\half{\frac12}

\def\d{\partial}

\def\inbar{\,\vrule height1.5ex width.4pt depth0pt}
\def\IC{\relax\hbox{$\inbar\kern-.3em{\rm C}$}}
\def\IR{\relax{\rm I\kern-.18em R}}
\def\IP{\relax{\rm I\kern-.18em P}}
\def\IZ{\relax{\rm I\kern-.18em Z}}

%
%
\def\np#1#2#3{Nucl. Phys. {\bf B#1} (#2) #3}

\catcode`\@=11
\def\slash#1{\mathord{\mathpalette\c@ncel{#1}}}
\overfullrule=0pt

\def\NN{{\cal N}}

\def\RR{{\cal R}}
\def\SS{{\cal S}}

\def\underrel#1\over#2{\mathrel{\mathop{\kern\z@#1}\limits_{#2}}}

\catcode`\@=12


%

\def \sinh{{\rm sinh}}
\def \cosh{{\rm cosh}}


\def\ls{{\bf L}}

\Title{ \rightline{hep-th/0412294}} {\vbox{\centerline{Strings in
a 2-d Extremal Black Hole}}}
\medskip
\centerline{\it Amit Giveon and Amit Sever}
\bigskip
\centerline{Racah Institute of Physics, The Hebrew University}
\centerline{Jerusalem 91904, Israel}
\centerline{giveon@vms.huji.ac.il, asever@phys.huji.ac.il}

\lref\gkcon{A. Giveon and D. Kutasov, unpublished.}

\bigskip\bigskip\bigskip
\noindent
String theory on 2-d charged black holes corresponding to
${SL(2)\times U(1)_L\over U(1)}$ exact asymmetric quotient CFTs
are investigated.
These backgrounds can be embedded, in particular, in a two dimensional
heterotic string. In the extremal case, the quotient CFT description
captures the near horizon physics, and is equivalent to strings in
$AdS_2$ with a gauge field. Such string vacua possess an infinite space-time
Virasoro symmetry, and hence enhancement of global space-time
Lie symmetries to affine symmetries, in agreement with the conjectured
$AdS_2/CFT_1$ correspondence. We argue that the entropy of these 2-d
black holes in string theory is compatible with semi-classical results,
and show that in perturbative computations part of an
incoming flux is absorbed by the black hole.
Moreover, on the way we find evidence that the 2-d heterotic string is
closely related to the $N=(2,1)$ string, and conjecture that they are dual
\gkcon.

\vfill

\Date{07/04}

\newsec{Introduction}

\lref\grs{
A.~Giveon, E.~Rabinovici and A.~Sever,
``Beyond the singularity of the 2-D charged black hole,''
JHEP {\bf 0307}, 055 (2003)
[arXiv:hep-th/0305140].
}

\lref\gkrs{
A.~Giveon, A.~Konechny, E.~Rabinovici and A.~Sever,
``On thermodynamical properties of some coset CFT backgrounds,''
arXiv:hep-th/0406131.
}

\lref\adsrev{
O.~Aharony, S.~S.~Gubser, J.~M.~Maldacena, H.~Ooguri and Y.~Oz,
``Large N field theories, string theory and gravity,''
Phys.\ Rept.\  {\bf 323}, 183 (2000)
[arXiv:hep-th/9905111].
}

\lref\gk{
A.~Giveon and D.~Kutasov,
``Little string theory in a double scaling limit,''
JHEP {\bf 9910}, 034 (1999)
[arXiv:hep-th/9909110].
}

\lref\gks{
A.~Giveon, D.~Kutasov and N.~Seiberg,
``Comments on string theory on AdS(3),''
Adv.\ Theor.\ Math.\ Phys.\  {\bf 2}, 733 (1998)
[arXiv:hep-th/9806194].
}

\lref\ls{
D.~A.~Lowe and A.~Strominger,
``Exact four-dimensional dyonic black holes and Bertotti-Robinson space-times
in string theory,''
Phys.\ Rev.\ Lett.\  {\bf 73}, 1468 (1994)
[arXiv:hep-th/9403186].
}

\lref\strominger{
A.~Strominger,
``AdS(2) quantum gravity and string theory,''
JHEP {\bf 9901}, 007 (1999)
[arXiv:hep-th/9809027].
}

\lref\btz{
M.~Banados, C.~Teitelboim and J.~Zanelli,
``The Black hole in three-dimensional space-time,''
Phys.\ Rev.\ Lett.\  {\bf 69}, 1849 (1992)
[arXiv:hep-th/9204099].
}

\lref\hw{
G.~T.~Horowitz and D.~L.~Welch,
``Exact three-dimensional black holes in string theory,''
Phys.\ Rev.\ Lett.\  {\bf 71}, 328 (1993)
[arXiv:hep-th/9302126].
}

\lref\kaloper{
N.~Kaloper,
``Miens of the three-dimensional black hole,''
Phys.\ Rev.\ D {\bf 48}, 2598 (1993)
[arXiv:hep-th/9303007].
}

\lref\hkv{
S.~Hemming and E.~Keski-Vakkuri,
``The spectrum of strings on BTZ black holes and spectral flow in the  SL(2,R)
WZW model,''
Nucl.\ Phys.\ B {\bf 626}, 363 (2002)
[arXiv:hep-th/0110252].
}

\lref\np{
C.~R.~Nappi and A.~Pasquinucci,
``Thermodynamics of two-dimensional black holes,''
Mod.\ Phys.\ Lett.\ A {\bf 7}, 3337 (1992)
[arXiv:gr-qc/9208002].
}

\lref\strom{
A.~Strominger,
``Black hole entropy from near-horizon microstates,''
JHEP {\bf 9802}, 009 (1998)
[arXiv:hep-th/9712251].
}

\lref\ao{
A.~Achucarro and M.~E.~Ortiz,
``Relating black holes in two-dimensions and three-dimensions,''
Phys.\ Rev.\ D {\bf 48}, 3600 (1993)
[arXiv:hep-th/9304068].
}

\lref\brhe{
J.~D.~Brown and M.~Henneaux,
``Central Charges In The Canonical Realization Of Asymptotic Symmetries: An
Example From Three-Dimensional Gravity,''
Commun.\ Math.\ Phys.\  {\bf 104}, 207 (1986).
}

\lref\gr{
A.~Giveon and M.~Rocek,
``Supersymmetric string vacua on AdS(3) x N,''
JHEP {\bf 9904}, 019 (1999)
[arXiv:hep-th/9904024].
}

\lref\bl{
D.~Berenstein and R.~G.~Leigh,
``Spacetime supersymmetry in AdS(3) backgrounds,''
Phys.\ Lett.\ B {\bf 458}, 297 (1999)
[arXiv:hep-th/9904040].
}

\lref\ags{
R.~Argurio, A.~Giveon and A.~Shomer,
``Superstrings on AdS(3) and symmetric products,''
JHEP {\bf 0012}, 003 (2000)
[arXiv:hep-th/0009242].
}

\lref\agss{
R.~Argurio, A.~Giveon and A.~Shomer,
``String theory on AdS(3) and symmetric products,''
Fortsch.\ Phys.\  {\bf 49}, 409 (2001)
[arXiv:hep-th/0012117].
}

\lref\aaaa{
A.~Giveon, A.~Konechny, A.~Pakman and A.~Sever,
``Type 0 strings in a 2-d black hole,''
JHEP {\bf 0310}, 025 (2003)
[arXiv:hep-th/0309056].
}

\lref\gkos{
A.~Giveon, B.~Kol, A.~Ori and A.~Sever,
``On the resolution of the time-like singularities in Reissner-Nordstroem and
negative-mass Schwarzschild,''
arXiv:hep-th/0401209.
}

\lref\giveon{
A.~Giveon,
``Target space duality and stringy black holes,''
Mod.\ Phys.\ Lett.\ A {\bf 6}, 2843 (1991).
}

\lref\gpr{
A.~Giveon, M.~Porrati and E.~Rabinovici,
``Target space duality in string theory,''
Phys.\ Rept.\  {\bf 244}, 77 (1994)
[arXiv:hep-th/9401139].
}

\lref\gsw{
M.~B.~Green, J.~H.~Schwarz and E.~Witten,
``Superstring Theory. Vol. 1: Introduction;''
``Superstring Theory. Vol. 2: Loop Amplitudes, Anomalies And Phenomenology.''
}

\lref\pol{
J.~Polchinski,
``String theory. Vol. 1: An introduction to the bosonic string;''
``String theory. Vol. 2: Superstring theory and beyond.''
}

\lref\maldacena{
J.~M.~Maldacena,
``The large N limit of superconformal field theories and supergravity,''
Adv.\ Theor.\ Math.\ Phys.\  {\bf 2}, 231 (1998)
[Int.\ J.\ Theor.\ Phys.\  {\bf 38}, 1113 (1999)]
[arXiv:hep-th/9711200].
}

\lref\gp{
A.~Giveon and A.~Pakman,
``More on superstrings in AdS(3) x N,''
JHEP {\bf 0303}, 056 (2003)
[arXiv:hep-th/0302217].
}

\lref\egkr{
S.~Elitzur, A.~Giveon, D.~Kutasov and E.~Rabinovici,
``From big bang to big crunch and beyond,''
JHEP {\bf 0206}, 017 (2002)
[arXiv:hep-th/0204189].
}

\lref\sads{
A.~Strominger,
``A matrix model for AdS(2),''
JHEP {\bf 0403}, 066 (2004)
[arXiv:hep-th/0312194].
}

\lref\gt{
G.~W.~Gibbons and P.~K.~Townsend,
``Black holes and Calogero models,''
Phys.\ Lett.\ B {\bf 454}, 187 (1999)
[arXiv:hep-th/9812034].
}

\lref\verlinde{
H.~Verlinde,
``Superstrings on AdS(2) and superconformal matrix quantum mechanics,''
arXiv:hep-th/0403024.
}

\lref\mny{
M.~D.~McGuigan, C.~R.~Nappi and S.~A.~Yost,
``Charged black holes in two-dimensional string theory,''
Nucl.\ Phys.\ B {\bf 375}, 421 (1992)
[arXiv:hep-th/9111038].
}

\lref\pierce{
D.~M.~Pierce,
``A (1,2) Heterotic String with Gauge Symmetry,''
Phys.\ Rev.\ D {\bf 53}, 7197 (1996)
[arXiv:hep-th/9601125].
}

\lref\km{
D.~Kutasov and E.~J.~Martinec,
``New Principles for String/Membrane Unification,''
Nucl.\ Phys.\ B {\bf 477}, 652 (1996)
[arXiv:hep-th/9602049].
}

\lref\fubini{
V.~de Alfaro, S.~Fubini and G.~Furlan,
``Conformal Invariance In Quantum Mechanics,''
Nuovo Cim.\ A {\bf 34}, 569 (1976).
}

\lref\bv{
E.~Bergshoeff and M.~A.~Vasiliev,
``The Calogero model and the Virasoro symmetry,''
Int.\ J.\ Mod.\ Phys.\ A {\bf 10}, 3477 (1995)
[arXiv:hep-th/9411093].
}

\lref\jjp{
I.~Jack, D.~R.~T.~Jones and J.~Panvel,
``Exact bosonic and supersymmetric string black hole solutions,''
Nucl.\ Phys.\ B {\bf 393}, 95 (1993)
[arXiv:hep-th/9201039].
}

\lref\bs{
I.~Bars and K.~Sfetsos,
``Conformally exact metric and dilaton in string theory on curved space-time,''
Phys.\ Rev.\ D {\bf 46}, 4510 (1992)
[arXiv:hep-th/9206006].
}

\lref\sfe{
K.~Sfetsos,
``Conformally exact results for SL(2,R) x SO(1,1)(d-2) / SO(1,1) coset
models,''
Nucl.\ Phys.\ B {\bf 389}, 424 (1993)
[arXiv:hep-th/9206048].
}

\lref\bsfe{
I.~Bars and K.~Sfetsos,
``SL(2-R) x SU(2) / IR**2 string model in curved space-time and exact conformal
results,''
Phys.\ Lett.\ B {\bf 301}, 183 (1993)
[arXiv:hep-th/9208001].
}

\lref\grt{
A.~Giveon, E.~Rabinovici and A.~A.~Tseytlin,
``Heterotic string solutions and coset conformal field theories,''
Nucl.\ Phys.\ B {\bf 409}, 339 (1993)
[arXiv:hep-th/9304155].
}

\lref\tseytlin{
A.~A.~Tseytlin,
``Conformal sigma models corresponding to gauged Wess-Zumino-Witten theories,''
Nucl.\ Phys.\ B {\bf 411}, 509 (1994)
[arXiv:hep-th/9302083].
}

\lref\egrsv{
S.~Elitzur, A.~Giveon, E.~Rabinovici, A.~Schwimmer and G.~Veneziano,
``Remarks on nonAbelian duality,''
Nucl.\ Phys.\ B {\bf 435}, 147 (1995)
[arXiv:hep-th/9409011].
}

\lref\olrd{
M.~J.~O'Loughlin and S.~Randjbar-Daemi,
``AdS(3) x R as a target space for the (2,1) string theory,''
Nucl.\ Phys.\ B {\bf 543}, 170 (1999)
[arXiv:hep-th/9807208].
}

\lref\grocek{
A.~Giveon and M.~Rocek,
``Generalized duality in curved string backgrounds,''
Nucl.\ Phys.\ B {\bf 380}, 128 (1992)
[arXiv:hep-th/9112070].
}

\lref\cj{
C.~V.~Johnson,
``Exact models of extremal dyonic 4-D black hole solutions of heterotic string
theory,''
Phys.\ Rev.\ D {\bf 50}, 4032 (1994)
[arXiv:hep-th/9403192].
}

\lref\dantwo{ D.~Israel, C.~Kounnas, D.~Orlando and
P.~M.~Petropoulos, ``Heterotic strings on homogeneous spaces,''
arXiv:hep-th/0412220.
}

In this note we consider string theory on a two dimensional
Extremal Black Hole (EBH), corresponding to the exact maximally
asymmetric quotient Conformal Field Theory (CFT) background
${SL(2,\IR)\times U(1)_L\over U(1)}$.
Strings propagating in 2-d black holes
may serve as useful toy models for the study of black holes
in a theory of quantum gravity (see \gpr\ for a review).
Our interest in this particular 2-d EBH is four fold:

\item{(1)}
It is an exact CFT background with a {\it constant} dilaton,
which thus can be used to study some aspects
of the black hole physics in standard perturbative string theory.

\item{(2)}
It has an infinite space-time Virasoro symmetry which may allow
a better understanding of its properties, such as the entropy.

\item{(3)}
It can be embedded in a {\it two dimensional} heterotic string theory,
which might have a useful Conformal Quantum Mechanics dual.

\item{(4)}
It is equivalent to string theory in $AdS_2$ with a gauge field,
and hence might shed more light on the conjectured
$AdS_2/CFT_1$ correspondence or, alternatively, be analyzed non-perturbatively
by using this duality.

We begin in section 2 by reviewing the 2-d Charged Black Holes
corresponding to a family of exact CFTs -- the (left-right asymmetric)
quotients ${SL(2,\IR)\times U(1)\over U(1)}$
\refs{\mny,\grocek,\gpr,\cj,\grs,\gkrs}.
In section 3 we consider the extremal case, and show that its corresponding
exact CFT -- the maximally asymmetric quotient -- captures the near horizon
physics, and is equivalent to an $AdS_2$ background with a gauge field.
In section 4 we sketch the construction of various string vacua
on $EBH\times\NN$, and in section 5 we construct a two dimensional
heterotic string on $EBH\equiv AdS_2$.
We show that in all string vacua on EBH there is a space-time Virasoro
symmetry, similar to string theory on $AdS_3$ \gks, which enhance some
global Lie symmetries to affine symmetries in the conjectured
space-time dual Quantum Mechanics.
On the way, we find evidence that the 2-d heterotic string is
closely related to the $N=(2,1)$ string \gkcon.
In section 6 we discuss the entropy of these 2-d black holes,
evaluated by using some aspects of its non-perturbative dual,
and recall that the reflection coefficient for waves scattered from
its event horizon, evaluated perturbatively in its exact CFT description,
is smaller than one.
Finally, in appendix A we describe $AdS_2$ in various coordinate systems,
and in appendix B we show that the EBH background is equivalent to
a maximally asymmetric orbifold of $AdS_3$ -- the extremal BTZ black hole.

\newsec{2-d Charged Black Holes as exact CFTs -- a review}

We consider string theory on the background $CBH\times \NN$, where
CBH is the 2-d Charged Black Hole background (described below) and $\NN$
is a compact Conformal Field Theory background.
The CBH background is the exact
${SL(2,\IR)_k\times U(1)\over U(1)}$ quotient CFT background
obtained by the anomaly free asymmetric gauging
\eqn\gauging{(g,x_L,x_R)\simeq\left(e^{\tau\cos(\psi)\sigma_3/\sqrt{k}}g
e^{\tau\sigma_3/\sqrt{k}}, x_L+\tau\sin(\psi), x_R\right)~.}
Here $g\in SL(2,\IR)_k$ -- an $SL(2,\IR)$ WZW model at level $k$ --
$(x_L,x_R)\in U(1)_L\times U(1)_R$, where $L,R$ stand for left- and
right-movers, respectively.
Note that the gauging \gauging\ does not act on $x_R$ and, therefore, this
background can be embedded, in particular,
in a 2-d heterotic string, where $x_R$ does not exist and with $x_L$
being part of its chiral internal space (see section 5).

The sigma-model background corresponding to this exact
${SL(2,\IR)_k\times U(1)\over U(1)}$ CFT is three dimensional.
Taking a small $U(1)$ radius (relative to the $SL(2,\IR)$ radius $\sqrt{k}$),
a Kaluza-Klein (KK) reduction to two dimensions with a gauge field is
justified. In a 2-d heterotic string this background is
${SL(2,\IR)_k\times U(1)_L\over U(1)}$, which is two dimensional
with a gauge field.

To obtain the sigma-model background, we first choose to parametrize
\eqn\param{g(z,y,\theta;\delta)=e^{(z+y)\sigma_3/2}g_{\delta}(\theta)
e^{(z-y)\sigma_3/2}~.}
For each choice of $g_{\delta}(\theta)$ this parametrization covers a certain
region of the $SL(2,\IR)$ group manifold (or its universal cover --
$\tilde{SL}(2,\IR)\equiv AdS_3$).
We denote by regions $A,B,C$ the ones corresponding to
\eqn\abc{g_A(\theta)=e^{\theta\sigma_1}~, \qquad
g_B(\theta)=e^{i\theta\sigma_2}~, \qquad
g_C(\theta)=e^{\theta\sigma_1}i\sigma_2~,}
respectively.
After gauging \gauging\ (with the gauge fixing choice $z=0$) and a KK
reduction one obtains, in each region,
a two dimensional metric, a dilaton $\Phi$ and a gauge field $A_y$
(for details see \refs{\grs,\gkrs}):~\foot{We set $l_s=1$.}
\eqn\aaa{\eqalign{A:\qquad {1\over k}ds^2=&d\theta^2-{\coth^2(\theta)\over
(\coth^2(\theta)-p^2)^2}dy^2~,\cr
{1\over\sqrt{k}}A_y=&{p\over p^2-\coth^2(\theta)}~,\qquad \theta\geq 0~,\cr
\Phi=&\Phi_0-{1\over 2}\log\left(\cosh^2(\theta)-p^2\sinh^2(\theta)\right)~,
}}
\eqn\bbb{\eqalign{B:\qquad {1\over k}ds^2=&-d\theta^2+{\cot^2(\theta)\over
(\cot^2(\theta)+p^2)^2}dy^2~,\cr
{1\over\sqrt{k}}A_y=&{p\over p^2+\cot^2(\theta)}~, \qquad
0\leq\theta\leq {\pi\over 2}~,\cr
\Phi=&\Phi_0-{1\over 2}\log\left(\cos^2(\theta)+p^2\sin^2(\theta)\right)~,
}}
\eqn\ccc{\eqalign{C:\qquad {1\over k}ds^2=&d\theta^2-{\tanh^2(\theta)\over
(\tanh^2(\theta)-p^2)^2}dy^2~,\cr
{1\over\sqrt{k}}A_y=&{p\over p^2-\tanh^2(\theta)}~,
\qquad \theta\geq 0~, \cr
\Phi=&\Phi_0-{1\over 2}\log\left(\sinh^2(\theta)-p^2\cosh^2(\theta)\right)~.
}}
Here,
\eqn\pp{p^2=\tan^2(\psi/2)~, \qquad \psi\in [0,\pi/2]~,}
where w.l.g. we chose $p^2\leq 1$, and $\Phi_0$ is a constant, related to
the dilaton on the group $SL(2,\IR)$ theory by
\eqn\dil{\Phi_{sl(2)}=\Phi_0-{1\over 2}\log\left({1+p^2\over 2}\right)~.}
Region A is obtained from region B by taking $\theta\to i\theta$, while
region C is obtained from B by $\theta\to i\theta +\pi/2$ \grs.

Next we do a coordinate transformation
\eqn\coortra{t={2y\over 1-p^2}~, \qquad r={2e^{-2\Phi}\over\sqrt{k}}~.}
In these Schwarzschild-like coordinates, the background \aaa\ - \ccc\
takes the form
\eqn\sch{{4\over k}ds^2=-f(r)dt^2+{dr^2\over r^2f(r)}~,\qquad
{1\over\sqrt k}A_t={Q\over 2}\left({1\over r_+} - {1\over r}\right)~,}
where
\eqn\fr{f(r)=1-{2M\over r}+{Q^2\over r^2}~,}
with
\eqn\mq{\sqrt{k}M=(1+p^2)e^{-2\Phi_0}~, \qquad \sqrt{k}Q=2pe^{-2\Phi_0}~,}
and
\eqn\inout{r_{\pm}=M\pm\sqrt{M^2-Q^2}~.}
Note that
\eqn\pprr{p^2={r_-\over r_+}~,}
where $p^2$ is defined in eq. \pp.

This background is a 2-d CBH with mass $M$ and charge $Q$ -- the one
discovered and studied in \mny.
The inner (Cauchy) and outer (event) horizons are located at $r=r_{\pm}$,
and the singularity is located at $r=0$.
Region A \aaa\ is mapped to the region outside the event horizon
$r_+<r<\infty$.
Region B \bbb\ is mapped to the region between the outer and inner horizons
$r_-<r<r_+$. Region C is mapped to the region beyond the inner horizon
$r<r_-$; the region beyond the singularity corresponds to $r<0$ or,
equivalently, a positive $r$ but $M<0$.~\foot{In
string theory, one may consider also the region beyond the singularity
\refs{\giveon,\gpr,\grs,\gkos}.}
The maximal extension of the CBH is given by taking the infinite regions
obtained from gauging the universal cover of $SL(2,\IR)$;
see \refs{\grs,\gkrs} for details.

In type II and type 0 string theory on $CBH\times\NN$ the CBH part
is a SCFT on ${SL(2,\IR)\times U(1)\over U(1)}$. The central charge
of the CBH superconformal sigma model is
$c_{cbh}=(3+6/k)+3/2$ and $c_{\NN}=15-c_{cbh}$, both for the left- and
right-movers. In the heterotic string, for which $x_R$ does not exist,
$c_{cbh}=3+6/k$ and $c_\NN=15-c_{cbh}$ for the fermionic
right-moving sector, while ${c_\NN}=26-c_{cbh}$ for the bosonic left-moving
sector once we embed the SCFT on CBH in the bosonic sector
and include the $U(1)_L$ connected to the EBH gauge field in $\NN_L$.
Finally, in the bosonic string $c_{cbh}=3+6/k$, with $k=k_{bosonic}-2$,
and ${c_\NN}=26-c_{cbh}$ both for the left- and right-movers.

In string theory, generically the exact coset CFT backgrounds corresponding
to the CBH receive higher order $\alpha'\sim 1/k$ corrections
(see e.g. \refs{\tseytlin,\grt} and references therein).
In string vacua corresponding to ``left-right symmetric''
supersymmetric quotients
we do not expect such corrections to the metric
\refs{\jjp,\bs,\sfe,\bsfe,\tseytlin,\grt}.~\foot{For
the 2-d heterotic string in such quotients --
where the background is supplemented by a (higher order in $\alpha'$)
gauge field which is set equal to the Lorentz connection  --
the metric does not receive $\alpha'$ corrections; see \grt\ for details.
}

\newsec{2-d Extremal Black Holes as exact CFTs}

\lref\dan{ D.~Israel, C.~Kounnas, D.~Orlando and
P.~M.~Petropoulos, ``Electric / magnetic deformations of S**3 and
AdS(3), and geometric cosets,'' arXiv:hep-th/0405213.}

\lref\gkir{A.~Giveon and E.~Kiritsis, ``Axial vector duality as a
gauge symmetry and topology change in string theory,'' Nucl.\
Phys.\ B {\bf 411}, 487 (1994) [arXiv:hep-th/9303016].}

We now consider string theory on the background $EBH\times\NN$,
where EBH is the 2-d Extremal Black Hole. This is, apparently, a
special case of the discussion in the previous section for which
$M^2=Q^2$ or, equivalently, $r_+=r_-$ $(p^2=1)$. In this case
$\psi=\pi/2$ \pp\ and hence the gauging \gauging\ acts only on
$SL(2,\IR)_R\times U(1)_L$:
\eqn\egau{(g,x_L,x_R)\simeq\left(ge^{\tau\sigma_3/\sqrt{k}},
x_L+\tau, x_R\right)~.} In the ``group coordinates'' $y,\theta$
or, for convenience, \eqn\groupcoor{\chi=2\theta~,} we find the
following metric and gauge field in the various regions. The 2-d
metric in regions A and C \aaa,\ccc\ is now identical, while the
gauge field is the same up to a sign (which can be changed by
$y\to -y$) and an additive constant, \eqn\aacc{A,C:\qquad {4\over
k}ds^2=d\chi^2-\sinh^2(\chi) dy^2~, \qquad {1\over\sqrt
k}A_y=\half(1\mp\cosh(\chi))~,} where the ($\mp$) in the gauge
field is ($-$) in region A and ($+$) in C. In region B \bbb, we
have \eqn\bb{B:\qquad {4\over k}ds^2=-d\chi^2+\sin^2(\chi) dy^2~,
\qquad {1\over\sqrt k}A_y=\half(1-\cos(\chi))~.} The dilaton is
constant, and equals in all the regions \aaa,\bbb,\ccc\ to its
value in the group \dil: \eqn\dilconst{\Phi=\Phi_0=\Phi_{sl(2)}~.}
This background is the same one obtained in \refs{\ls,\strominger}
by a certain compactification of $AdS_3$ on a circle, and it is
equivalent to a KK reduction of the extremal BTZ black hole to two
dimensions; see appendix B for details. This background and its
properties are also obtained by a certain current-current
deformation of $SL(2,\IR)\times U(1)$ in \dan, along the lines of
\gkir.

Each of the backgrounds \aacc\ and \bb\ covers a different patch of $AdS_2$
(see appendix A for more details on $AdS_2$).
To show this, we first represent $AdS_2$ (with a unit radius, for simplicity)
as the infinite cover of the hyperboloid
\eqn\hyperbol{X_0^2+X_2^2-X_1^2=1~,}
in $\IR^{2,1}$ with metric
\eqn\metfla{ds^2=-dX_0^2-dX_2^2+dX_1^2~.}
Parametrizing
\eqn\parads{\eqalign{X_0=&\sinh(\chi)\sinh(y)~,\cr
            X_1=&\sinh(\chi)\cosh(y)~,\cr
            X_2=&\cosh(\chi)~,}}
one obtains the metric \aacc.
Hence, region A (or C) covers a patch of $AdS_2$ with $X_2>1$ and $X_1>0$
(or $X_2<-1$ and $X_1<0$~\foot{Recall
that region C is obtained from A by $\chi\to\chi-i\pi$
(see eq. (26) in \grs).}).
Similarly, parametrizing
\eqn\paradsb{\eqalign{X_0=&\sin(\chi)\cosh(y)~,\cr
            X_1=&\sin(\chi)\sinh(y)~,\cr
            X_2=&\cos(\chi)~,}}
one obtains the metric \bb. Hence, region B covers the patch $-1<X_2<1$
and $X_0>0$ of $AdS_2$.

To get the maximal extension of the EBH, one considers the infinite regions
obtained from gauging the universal cover of $SL(2,\IR)$
(see \refs{\grs,\gkrs} for details).
This maximally extended EBH covers the full $AdS_2$ --
the universal cover of the hyperboloid \hyperbol\ (see appendix A).
The location of the various regions in the $AdS_2$ strip is
described in figure 1.

 \fig{The regions of the Extremal Black Hole in $AdS_2$; $\tau,\varphi$ are
 the coordinates of section A.2 and $A,B,C \leftrightarrow A',B',C'$ under
 $\chi \leftrightarrow \chi'$ in \parads,\paradsb.}{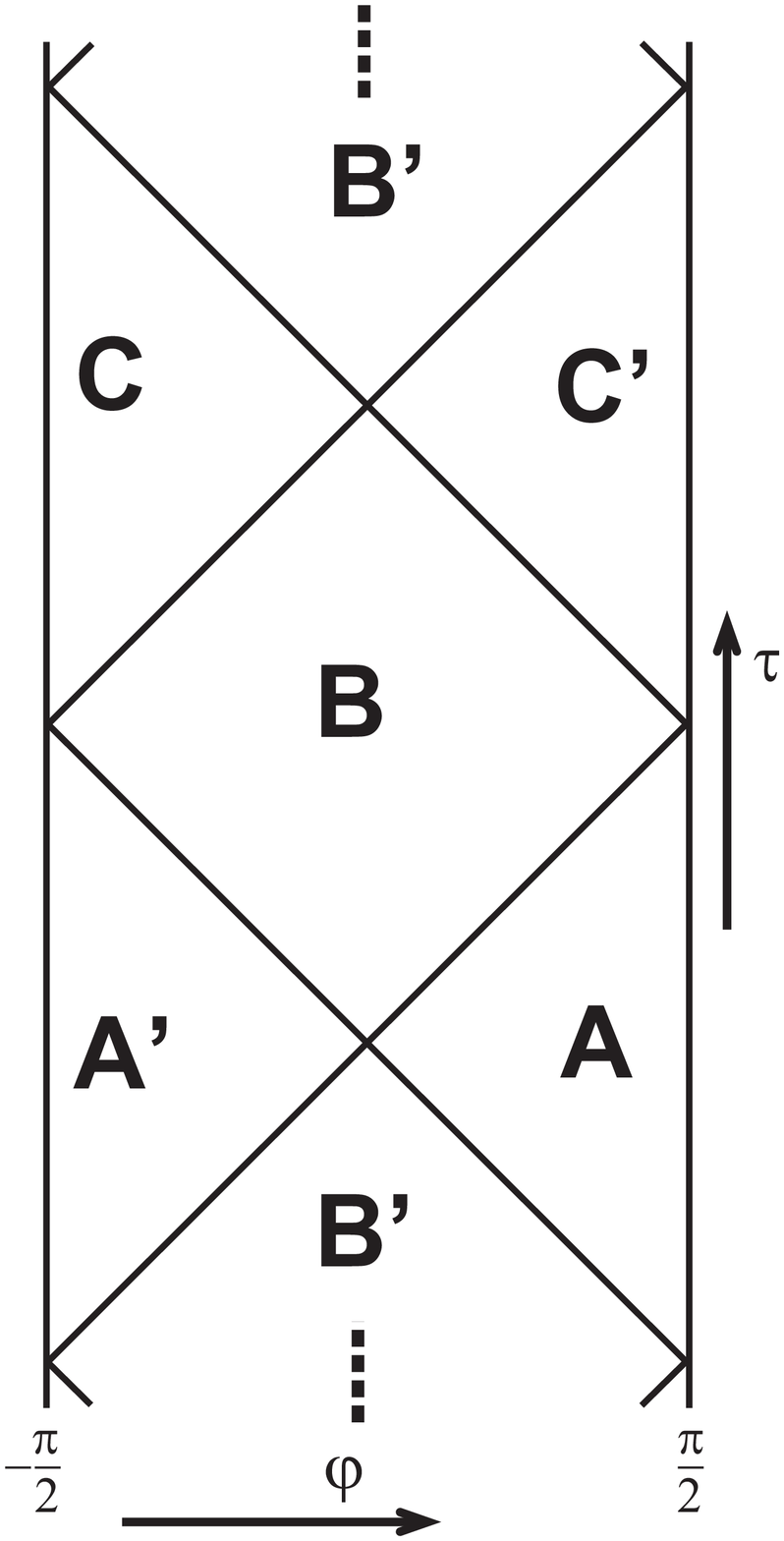}{6 truecm}
 \figlabel\ebhh

What happens if instead we take $M^2=Q^2$ in the 2-d background written
in the Schwarzschild-like coordinates \sch,\fr?
As in the non-extremal case, one finds an asymptotically flat background
with a linear dilaton, namely, at $r\to\infty$:
\eqn\asflat{ds^2\to {k\over 4} (-dt^2+d\phi^2)~,
\qquad \Phi\to -{1\over 2}\phi+const~, \qquad A_t\to const~, \qquad
r=e^{\phi}~.}
This background is thus different from the $AdS_2$ background above.
The reason for this is that the coordinate transformation \coortra\
is ill defined in the extremal case: the dilaton is a constant \dilconst\
and hence $r$ in eq. \coortra\ is fixed at the horizon \inout,\mq: $r=M$.
Consequently,
to compare to the quotient CFT background in the $p=1$ case, we are
led to investigate the near horizon limit of the extremal 2-d black hole.

The near horizon limit is obtained, for instance, by inserting $r=M+u$ in
\sch\ with $f(r)=(1-M/r)^2$, and keeping only terms with $u\ll M$.
After rescaling $t$, one obtains
\eqn\dsads{{4\over k}ds^2=-u^2dt^2+{du^2\over u^2}~, \qquad
{1\over\sqrt k}A_t={u\over 2}~.}
For $u\geq 0$, this background is the Poincar\'e patch of $AdS_2$ -- it covers
one half of the hyperboloid \hyperbol\ (see appendix A and
\adsrev\ for a review) -- with a gauge field.

We have thus learned that the exact CFT background
${SL(2,\IR)\times U(1)\over U(1)}$ with the gauged $U(1)$ acting
in a maximally asymmetric way \egau\
describes both the physics of string theory in the near horizon limit of an
extremal 2-d black hole or, equivalently, in an $AdS_2$
background with a gauge field.
Another relation between \aacc,\bb\ and an $AdS_2$ black hole is described
in subsection A.4 of appendix A.

\newsec{String Theory on $EBH\times\NN$}

We now discuss various perturbative
string vacua (see \refs{\gsw,\pol} for a review)
in the EBH background. In subsection 4.1 we begin with the bosonic
string, in subsection 4.2 we sketch the construction of type II
and type 0 fermionic string vacua, and in subsection 4.3 the heterotic
string is introduced.

\subsec{Bosonic String Theory on $EBH\times\NN$}

The bosonic string theory has tachyons, but as usual it is
constructive to consider first the bosonic string on $EBH\times\NN$.
In this case $EBH\times\NN$ is a CFT background with $c=26$.
Physical vertex operators are obtained by dressing an operator $V_\NN$
in the CFT $\NN$ with an operator in $EBH$, such that the on-shell mass
condition and other physical conditions are satisfied.
For instance, some vertex operators in $EBH$ are obtained from
primaries of $SL(2,\IR)\times U(1)$,
\eqn\primbos{V_{jm\bar m}e^{i(k_Lx_L+k_Rx_R)}~,}
by imposing the gauge condition (see eq. (52) in \grs)
\eqn\gc{\bar m=\sqrt{k}k_L~.}
Here $V_{jm\bar m}$ is a primary of $SL(2,\IR)_L\times SL(2,\IR)_R$ with
Casimir $-j(j+1)$ and $(J^3,\bar J^3)$ eigenvalues equal to
$(m,\bar m)$,~\foot{We actually consider
the CFT on Euclidean $AdS_3$, for which $(m,\bar m)$
in eq. \primbos\ are real.
To obtain results in the Lorentzian
black hole we have to analytically continue $m\to im$
(see \refs{\grs,\aaaa} for more details).}
and $(k_L,k_R)$ are momenta on the even self-dual Narain lattice
$\Gamma^{1,1}$.
In particular, for uncharged states in EBH ($k_L=0$) the gauge condition
\gc\ reads $\bar m=0$.
Thus, while the right-moving $SL(2,\IR)_R$ symmetry is broken to $U(1)$
(only the generator $\bar L_0$ survives the gauge condition), the left-moving
$SL(2,\IR)_L$ survives \gc. Moreover, the global $SL(2,\IR)$ Lie algebra --
generated by $L_0, L_{\pm 1}$ --
is enhanced to a full Virasoro symmetry in target space,
generated by the operators $L_n$ with
$n\in {\bf Z}$, constructed in eq. (2.36) of \gks.

The $AdS_2$ nature of EBH suggests that this Virasoro symmetry
is the conformal group of a certain Conformal Quantum Mechanics,
as expected in the $AdS_2/CFT_1$ conjecture \maldacena\
(see \adsrev\ for a review).
For a closely related discussion see \strominger.

\subsec{Type II and 0 String Theory on $EBH\times\NN$}

To construct fermionic strings we consider
the SCFT on $EBH\times\NN$ with a central charge $c=15$.
If the CFT background $\NN$ has an affine $U(1)$ symmetry,
and $\NN/U(1)$ has an $N=2$ supersymmetry, the worldsheet theory
on $EBH\times\NN$ allows to construct string vacua with
space-time superconformal symmetries.
Such superstrings on $EBH\times\NN$
can be constructed in a similar way to the supersymmetric string vacua on
$AdS_3\times\NN$, studied in \refs{\gr,\bl,\gp}. Generically, only one half
of the superconformal symmetries in $AdS_3$ survive in the superstring on EBH
-- the ones associated with the $SL(2,\IR)_L$ left-moving symmetry,
which survives the gauge condition \gc.
Type 0 string theory on $EBH\times\NN$ can be obtained in a standard way.
Both type II and 0 string vacua on EBH have a space-time Virasoro
symmetry generated by the operators $L_n$ constructed in eq. (3.5) of \gks.
The construction of physical vertex operators in these theories is
a straightforward generalization of the discussion in $AdS_3\times\NN$
\refs{\gks,\ags,\agss} and the previous subsection, with the
gauge condition \gc, the appropriate mass shell condition, other physical
conditions and GSO projection applied.

\subsec{Heterotic String Theory on $EBH\times\NN$}

In the heterotic string on $EBH\times\NN$
we consider the superconformal extension of
of $EBH$, and embed it such that the right-moving SCFT is part of the
fermionic sector, while the left-moving SCFT is part of the bosonic sector.
Moreover, w.l.g. we choose $x_L$ in \egau\ to be part of the chiral internal
space of $\NN$, hence $x_R$ does not exist.
Consequently, the sigma model background obtained by the gauging \egau\ is
a {\it two dimensional} $AdS_2$ background with a gauge field, given in
\aacc,\bb.

As in the bosonic string, the left-moving Virasoro generators $L_n$
survive the gauge condition \gc. Hence the target space dual -- via the
conjectured $AdS_2/CFT_1$ correspondence -- is expected to be a certain
Conformal Quantum Mechanics with a full Virasoro algebra as its
conformal symmetry.

A particularly interesting case is the two dimensional heterotic string
on $EBH$ or, equivalently, the heterotic string on $AdS_2$.
This is the topic of the next section.

\newsec{2-d Heterotic String Theory on $EBH\equiv AdS_2$}

The structure of the CBH background -- ${SL(2,\IR)\times U(1)_L\over U(1)}$ --
allows us to embed it in a two dimensional heterotic string theory.
To discuss the 2-d heterotic string
on EBH (or, equivalently, on $AdS_2$ with a gauge field),
we first consider the slightly simpler, closely related
case of the $(2,2)$ heterotic string on $SL(2,\IR)/U(1)$
-- the uncharged 2-d black hole.

The right-moving fermionic sector of this heterotic string consists entirely
of the right-moving sector of the SCFT on $SL(2,\IR)/U(1)$, hence the
$SL(2,\IR)$
level is $k=1/2$, such that $c_{fermionic}=3+6/k=15$ is critical.
The SCFT on $SL(2,\IR)/U(1)$ has an $N=(2,2)$ supersymmetry. The right-moving
$U(1)_{\RR}$ current of this $N=2$ algebra is
$\bar J_{\RR}=i\sqrt 5\bar\partial\bar H$,
where $\bar H(\bar z)$ is a canonically normalized scalar:
$\bar H(\bar z)\bar H(\bar w)\sim -\log(\bar z -\bar w)$.
Space-time supercurrents correspond to chiral spin fields on the
worldsheet:
\eqn\stss{Q_{\pm}\equiv \int d\bar z e^{-{\bar\varphi\over 2}}
e^{\pm{i\over 2}{\sqrt 5}\bar H}~,}
where $\bar\varphi (\bar z)$ is the scalar appearing in
the bosonization of the superconformal ghosts.

The left-moving bosonic sector consists of
the same $N=2$ SCFT with central charge $c=15$,
and with a $U(1)$ $\RR$-current:
\eqn\jr{J_{\RR}=i\sqrt{c\over 3}\partial H=i\sqrt 5\partial H~,}
where $H(z)$ is a canonically normalized scalar.
In addition, there are $22$ free chiral fermions $\lambda^A(z)$,
$A=1,..,22$, such that the total central charge is critical:
$c_{bosonic}=15+\half\times 22=26$.

One can obtain consistent, stable 2-d heterotic string vacua by either
performing a diagonal GSO projection or a chiral GSO projection
-- mutual locality with \stss.~\foot{In the latter case it is not clear to
us if there is a consistent non-trivial theory; in the former case,
the consistency of the theory in its ``flat limit'' will be discussed below.}
Moreover, there are various ways to construct such modular invariant heterotic
vacua, each is specified by its gauge symmetry.
The latter is generically generated by left-moving currents.
We now discuss two consistent heterotic vacua -- those which
correspond to a symmetric embedding of the gauge connection in
the spin connection:

\item{(1)}
{\it $U(1)\times SO(22)$}:
We can construct a consistent heterotic string
background with ${22\times 21\over 2}$ holomorphic currents in the
adjoint representation of $SO(22)$:
\eqn\sod{\lambda^A\lambda^B~, \qquad A,B=1,..,22 \qquad
\in~Adjoint~of~SO(22)~.}
Together with the $U(1)_{\RR}$ current $i\partial H$ in \jr\ we have a
$U(1)\times SO(22)$ gauge symmetry.

\item{(2)}
{\it $SU(5)\times E_8$}:
This heterotic string has an $E_8$ gauge symmetry generated by
chiral currents in the adjoint of $SO(16)$, given by $120$ bilinears in
16 of the $\lambda$'s, say $\lambda^{7},..,\lambda^{22}$,
and the spinor $128$ of this $SO(16)$ generated by
the spin fields of these 16 $\lambda$'s.
In addition there is an $E_{4}\equiv SU(5)$ gauge
symmetry, whose holomorphic currents are:
\eqn\jdftz{i\partial H~,\quad \lambda^a\lambda^b~,~a,b=1,..,6~,\quad
e^{{i\over 2}\sqrt 5 H}S_\alpha~,\quad
e^{-{i\over 2}\sqrt 5 H}S_{\bar\alpha}~,}
where $S_\alpha$ and $S_{\bar\alpha}$ are spin fields in the
$4$ and $\bar 4$ representations of $SO(6)$, respectively.
Hence, the currents \jdftz\ are
in the adjoint ($1\oplus {6\times 5\over 2}$) plus spinor and spinor bar
($4\oplus \bar 4$) of $U(1)\times SO(6)$,
respectively, which is equivalent to the adjoint of $SU(5)$:
\eqn\rzero{1\oplus 15\oplus
4\oplus \bar{4}~~of~~U(1)\times SO(6)~=~24~~of~~SU(5)~.
}
We thus find that the gauge symmetry is $SU(5)\times E_8$.

In the ``flat limit'' $SL(2,\IR)/U(1)\to \IR_\phi\times S^1$ (or
$\IR_\phi\times \IR_t$ in the Lorentzian case), where $\IR_\phi$
is the SCFT of a scalar with a linear dilaton~\foot{To obtain a
weakly coupled string theory, we can re-add either an $N=1$
Liouville superpotential or an $N=2$ Liouville superpotential. The
latter case is dual to the $SL(2,\IR)/U(1)$ SCFT \gk. In the
former case only a diagonal GSO projection can be done -- the one
discussed below.} (and $\IR_t$ is time), the $U(1)\times SO(22)$
and $SU(5)\times E_8$ are enhanced to $SO(24)$ and $SO(8)\times
E_8$, respectively. The enhanced symmetry is due to the existence
of two extra free fermions -- $\psi_\phi$ and $\psi_t$ -- which
together with the $22$ $\lambda$'s now give rise to these enlarged
gauge groups. In this case, the one loop partition function of a
2-d heterotic string with a diagonal GSO projection was computed
in \mny. For the $SO(24)$ string it is the modular invariant
function \eqn\zso{Z_{SO(24)}(\tau)= {1\over
2\eta^{12}}\left(\theta_3^{12}-\theta_4^{12}-\theta_2^{12}\right)~,
} where $\theta_{2,3,4}$ are ``theta functions'' and $\eta$ is the
``Dedekind eta function'' (see e.g. \pol\ for a review on such
functions, and \mny\ and references therein for a detailed
computation of this partition function). Actually, this modular
invariant function is a constant $Z_{SO(24)}(\tau)=24$. Indeed,
the physical spectrum of the theory consists in this case of $24$
massless scalars in the fundamental representation of $SO(24)$, in
the (NS,NS) sector, whose corresponding vertex operators
are~\foot{For simplicity, here and below we write the vertex
operators corresponding to zero energy; for non-zero energy $E$ an
$e^{iEt}$ piece should be added and the Liouville dressing factor
should be changed accordingly to $e^{(-1+iE)\phi}$, so that the
on-shell mass condition is satisfied.}
\eqn\vso{V^I=e^{-\bar\varphi(\bar z)} \lambda^I(z)e^{-\phi(z,\bar
z)}~,\qquad I=1,...,24~,} where $\lambda^I$ are the $24$ free
fermions constituting the $SO(24)$ ``internal space.'' For the
$SO(8)\times E_8$ theory one finds \eqn\zsoe{Z_{SO(8)\times
E_8}(\tau)= {1\over
4\eta^{12}}\left(\theta_3^{4}-\theta_4^{4}-\theta_2^{4}\right)
\left(\theta_3^{8}+\theta_4^{8}+\theta_2^{8}\right)~.} Here
$Z_{SO(8)\times E_8}(\tau)=0$, and indeed there is an equal number
of space-time bosons and fermions:  $8$ massless scalars in the
fundamental representation $8_v$ of $SO(8)$, in the (NS,NS)
sector, whose vertex operators are
\eqn\vsoe{V^a=e^{-\bar\varphi(\bar z)} \lambda^a(z)e^{-\phi(z,\bar
z)}~,\qquad a=1,...,8~,} where $\lambda^a$ are the $8$ free
worldsheet fermions constituting the $SO(8)$ part of internal
space, and $8$ chiral and anti-chiral space-time fermions, in the
(R,R) sector~\foot{Note that in the heterotic string every
operator whose right-moving part is in the Ramond sector
corresponds to a space-time fermion.}, with vertex operators:
\eqn\chia{\chi_\alpha=e^{-{\bar\varphi\over 2}} e^{+{i\over 2}\bar
{\cal H}}\SS_\alpha e^{-\phi}~,\qquad
\bar\chi_{\bar\alpha}=e^{-{\bar\varphi\over 2}} e^{-{i\over 2}\bar
{\cal H}}\SS_{\bar\alpha} e^{-\phi}~,} where $\SS_\alpha$ and
$\SS_{\bar\alpha}$ are spin fields in the $8_s$ and $8_c$ spinor
representations of $SO(8)$, and $i\d\bar {\cal
H}=\bar\psi_t\bar\psi_\phi$.

Remarkably, as was observed more than a year ago \gkcon, the
partition functions \zso\ and \zsoe\ are identical to the
partition functions in the $N=(2,1)$ heterotic string, computed in
\pierce\ and as appear in eqs. (3.1) and (3.17) of \km,
respectively. Moreover, the physical spectrum in \vso\ and
\vsoe,\chia\ is identical to the one in the $N=(2,1)$ heterotic
string; there is a one to one correspondence with the vertex
operators in eqs. (3.2) and (3.20),(3.21) of \km, respectively.
There is however an important difference between the two theories:
the $N=(2,1)$ heterotic string in flat space has a $1+1$
dimensional Poincar\' e symmetry, while here the space-like
direction $\phi$ has a linear dilaton, with all its consequences.
Nevertheless, it is natural to expect that the 2-d heterotic
string is dual to the $N=(2,1)$ string on a different background.

We conjecture \gkcon\ that the 2-d heterotic string on $\IR_\phi\times\IR_t$
is dual to the $N=(2,1)$ string on
$\IR_\phi\times\IR_t\times\IR_x\times\IR_\rho$,
where $\IR_\rho$ is the SCFT of a time-like scalar with a linear dilaton
and $\IR_x$ is a space-like scalar~\foot{This SCFT on $\IR^{2,2}$ with a
null-like linear dilaton is the ``near horizon'' of
a solitonic configuration in $\IR^{2,2}$, of the sort studied in
\olrd.}.
Fixing the null Abelian gauge by setting $p_x=p_\rho=0$,
we obtain~\foot{The
gauging in a direction with a linear dilaton may be done along the lines
of eqs. (6.40) -- (6.42) in \egrsv. Gauging
the $\IR^{1,1}_{t,x}$ directions we would obtain, instead, an $\IR^{1,1}$
with a null-like linear dilaton.} in vertex
operators a factor $e^{iEt+(-1+iE)\phi}e^{-\rho}$
(instead of the factor $e^{iEt+iEx}$ in flat space).
Now the two theories have the same space-time symmetries.
It remains to show that such an $N=(2,1)$ string is consistent
and that its correlators are identical to those of the 2-d heterotic string.

We now turn to the two dimensional heterotic string on
$EBH\times\NN$. In this case the right-moving part of $\NN$ is
trivial, and hence the $SL(2,\IR)$ level is set to $k=1/2$, as
above, such that the left-moving central charge is critical
($3+6/k=15$). The left-moving bosonic sector consists of the SCFT
on EBH (with $c=15$) and an additional internal chiral sector
which can be constructed, for instance, by the same $22$ free
chiral fermions as above. Two out of these fermions correspond now
to the fermionization of $x_L$ -- the internal direction connected
to the gauge field on EBH. The global gauge symmetry is now
$SL(2,\IR)\times U(1)\times SO(20)$ or $SL(2,\IR)\times U(1)\times
SO(4)\times E_8$. The former is obtained as follows: $U(1)\times
SO(22)$ is broken by the gauge field to $U(1)^2\times SO(20)$, and
one of the $U(1)$'s is enhanced to an $SL(2,\IR)$. The latter is a
consequence of the breaking $SU(5)\to U(1)^2\times SO(4)$ by the
EBH background, with the enhancement of one $U(1)$ to an
$SL(2,\IR)$.~\foot{There are other ways to pick up a $U(1)$
subgroup -- with a level bigger than $1$ -- which will break
$SO(22)$ or $SU(5)$ differently \dantwo; we thank D. Israel for
pointing this out to us.} Since the $SL(2,\IR)$ global symmetry is
actually further enhanced to a full Virasoro algebra, the global
space-time Lie symmetry $G=U(1)\times SO(20)$ or $U(1)\times
SO(4)\times E_8$ is enhanced to affine $G$ in the dual $CFT_1$, as
in eq. (2.27) of \gks.

In a similar way, one can construct two dimensional heterotic strings on the
non-extremal 2-d CBH backgrounds \sch.
Two dimensional string theories in asymptotically flat space
(with a linear dilaton) are expected to be dual to certain large $N$
Matrix Quantum Mechanics theories (MQM), which can be described
by $N\to\infty$ free fermions in an inverted harmonic oscillator
potential. It is thus natural to conjecture that for the 2-d heterotic string
on the asymptotically flat 2-d non-extremal
CBH background \sch\ a similar duality holds
(perhaps along the lines of the proposal in section 6 of \aaaa\ for the
type 0 string theory on the uncharged 2-d black hole).
Such MQM should have a global symmetry $G$.

For the extremal black hole we have obtained, instead,
a two dimensional heterotic string on $AdS_2$.
In this case we expect a dual $CFT_1$. Such a $CFT_1$ can perhaps
be similar to the ones proposed for type 0A or type IIA string theory
on $AdS_2$:
a certain Conformal Quantum Mechanics, which might be related
to (a supersymmetric extension of) a theory of free fermions in a potential
of the form $V\simeq Q^2/r^2$ (as proposed in \sads\ for type 0A on $AdS_2$
with $Q$ units of RR flux), or perhaps a version of a supersymmetric
Calogero-Moser model (as proposed for type IIA on $AdS_2$ in
\refs{\gt,\verlinde}, where the above eigenvalue potential
$V\simeq \sum_{i=1}^n {Q^2\over r_i^2}$ is replaced by
$V\simeq \sum_{i<j}^n {Q^2\over (r_i-r_j)^2}$).

There is however a serious problem with these $CFT_1$ candidates.
The first candidate has an $SL(2,\IR)$ symmetry \fubini, but we do not know
how to enhance it to a Virasoro symmetry. The Calogero models have a Virasoro
and other affine symmetries \bv, but they both have zero central charges.
On the other hand, our space-time theories have a Virasoro symmetry with
central charge $c_{st}\simeq 1/g_s^2$ (see the next section) and affine
symmetries with level $k_{st}\simeq 1/g_s^2$. Hence, it seems that
the dual space-time theory has the properties of an {\it holomorphic} $CFT_2$
with a large central charge -- closely related to the symmetric product duals
to string theory on $AdS_3$, as follows from the facts in appendix B and the
study in \refs{\ags,\agss} --
rather than one of the known $CFT_1$'s discussed above.

\newsec{Entropy and Reflection}

In this section we discuss the entropy of the 2-d extremal black hole
evaluated by using some aspects of its non-perturbative space-time dual,
and the reflection coefficient of waves scattered from its event
horizon by using string perturbation theory in the exact CFT description.

Some thermodynamical properties of the 2-d CBH \sch\ were studied
semi-classically in \refs{\np,\gkrs}.
The entropy of the CBH is~\foot{We shall not
follow $(M,Q)$-independent (though $k$-dependent) factors in the entropy.}
\eqn\ent{S_{CBH}\simeq r_+~,}
where $r_+$ is given in \inout.
In particular, in the extremal case the entropy is
\eqn\eent{S_{EBH}\simeq M\simeq {1\over g_s^2}~,}
where in the last equality we used the dependence of the 2-d black hole mass
on the string coupling \mq.

In string theory on $EBH\times\NN$ it is interesting to understand
how to obtain such an entropy by counting the number of microstates.
Generically, the study in this work provides a partial answer.
Since the EBH background is equivalent to the extremal BTZ (see appendix B),
they share the same entropy. In string theory on the BTZ black hole,
the degrees of freedom consist of states in the Ramond sector
with $L_0+\bar L_0\simeq M_{BTZ}$ and  $L_0-\bar L_0\simeq J_{BTZ}$,
in a dual space-time $CFT_2$ \refs{\brhe,\strom,\gks}.
The semiclassical entropy of the BTZ black hole
indeed matches the number of states in this $CFT_2$,
as obtained from the Cardy formula \strom:
\eqn\sbtz{S_{BTZ}=2\pi\left(\sqrt{c_{st}L_0\over 6}
+\sqrt{c_{st}\bar L_0\over 6}\right)~,}
where $c_{st}$ is the central charge of the space-time CFT.
The KK reduction of the BTZ black hole to two dimensions
yields a 2-d charged black hole \refs{\ao,\ls}, whose near horizon behavior
is identical to the the near horizon of our 2-d CBH \sch, with
$M_{BTZ}\simeq M$ and $J_{BTZ}\simeq Q$
(see eq. (2.9) in \ls\ and compare
to \sch\ - \mq, in the near horizon).
Hence, the entropy in the extremal limit
behaves like $\sqrt{c_{st}M}$,~\foot{Note that although the KK reduction of
the BTZ black holes to 2-d yields asymptotically $AdS_2$ black holes
\refs{\ao,\ls}, unlike \sch\ - \mq\ which are asymptotically flat,
we expect the entropy of both black holes
to be dominated by near horizon microstates, and hence to be similar
even in the non-extremal case.}
and since $c_{st}\simeq 1/g_s^2$ \refs{\brhe,\strom,\gks}, we obtain
the same result as in eq. \eent.
We thus conclude that the same near horizon microstates constituting the
entropy of the extremal BTZ black hole, lead to an entropy of the 2-d EBH
which is linear in the black hole mass, as expected semiclassically.

Is eq. \eent\ valid also for the near horizon microstates in the
two dimensional heterotic string on EBH? Since the extremal BTZ has
$L_0\simeq M$ and $\bar L_0=0$,
it is plausible that the same space-time right-moving Ramond
ground states with left-moving excitations -- originated from
(non-perturbative)
string excitations and constituting the entropy of the BTZ black hole --
also give rise to the same
entropy $S\simeq M$ in the 2-d heterotic string on $EBH$.
If true, this result should be also compatible with the number of states in
the $CFT_1\equiv CMQM$ dual to the 2-d heterotic string on $EBH\equiv AdS_2$.


The reflection coefficient $R(p)$ of a wave scattered from the event horizon
of the 2-d CBH can be obtained in perturbative string theory
from the two point function of the exact CFT background,
by a formal analytic continuation,
as in \refs{\egkr,\grs,\aaaa}. For vertex operators of the type
\primbos\ one finds
\eqn\rrr{R(j;m,\bar m)={\Gamma(1-{2j+1\over k})\over \Gamma(1+{2j+1\over k})}
{\Gamma(-2j-1)\over \Gamma(2j+1)}
{\Gamma(1+j+im)\Gamma(1+j+i\bar m)\over\Gamma(-j+im)\Gamma(-j+i\bar m)}~,}
where $j$ is related to the momentum $p$ along the radial direction by
$j=-\half+ip$ and $(m,\bar m)$ are related to each other and to $p$
by the gauge condition and the mass shell condition; hence $m$ is related
to the energy.
As discussed in \grs, one obtains $|R(p)|<1$.
For instance, for uncharged waves in the extremal case, using the gauge
condition \gc\ in \rrr, one finds
\eqn\rlone{|R(p,m)|={\cosh(\pi(2p-m))+\cosh(\pi m)\over
\cosh(\pi(2p+m))+\cosh(\pi m)}~.}
For incoming waves with energy $m$ and momentum $p$ ($p,m>0$),
we thus obtain
\eqn\rlo{|R(p)|<1~.}
Therefore, part of the incoming flux is absorbed by the black hole.

\bigskip
\noindent{\bf Acknowledgements:} The relation of the 2-d heterotic
string to the $N=(2,1)$ string, and the conjecture that they are
dual, was done with David Kutasov \gkcon\ during A.G.'s visit at
the EFI and the Department of Physics at the University of Chicago
in September, 2003. We thank O.~Aharony, M.~Berkooz, D.~Israel and
D.~Kutasov for discussions. A.G. and A.S. thank the EFI and the
Department of Physics at the University of Chicago for its warm
hospitality. A.G. thanks the MPI for Physics in Munich for its
warm hospitality. This work was supported in part by the Israel
Academy of Sciences and Humanities -- Centers of Excellence
Program, by GIF -- German-Israel Bi-National Science Foundation,
and the European RTN network HPRN-CT-2000-00122. The work of A.S.
is supported in part by the Horowitz Foundation.

\appendix{A}{$AdS_2$}

Two dimensional Anti-de-Sitter space ($AdS_2$)
can be represented as (the infinite cover of)
hyperboloid (see \adsrev\ for a review)
\eqn\hyperbol{X_0^2+X_2^2-X_1^2=R^2~,}
in $\IR^{2,1}$ with metric
\eqn\metflat{ds^2=-dX_0^2-dX_2^2+dX_1^2~.}
There are various useful coordinate systems which we describe in the
following subsections.

\subsec{Global Coordinates}

A solution of \hyperbol\ is given, for instance, by parametrizing
\eqn\globalads{\eqalign{X_0=&R\cosh\rho\cos\tau~,\cr
            X_1=&R\sinh\rho~,\cr
            X_2=&R\cosh\rho\sin\tau~.}}
Using \metflat\ with \globalads, one obtaines the metric of $AdS_2$ as
\eqn\metglobal{ds^2=R^2(-\cosh^2\rho d\tau^2+d\rho^2)~.}
For $\rho\in\IR$ and $0\leq\tau\leq 2\pi$ the solution \globalads\
covers the whole hyperboloid \hyperbol\ exactly once. Hence, $(\tau,\rho)$
are called global coordinates of $AdS_2$.
Near $\rho=0$ the metric behaves as $ds^2\to R^2(- d\tau^2+d\rho^2)$, hence
the hyperboloid has the topology of $S^1\times \IR$, where $S^1$ represents
closed time-like curves in the $\tau$ direction.
By unwraping $\tau$ to $-\infty<\tau <\infty$ we obtain the infinite cover
of the hyperboloid, and avoid the problems of closed time-like curves.
Usually, by $AdS_2$ we mean this universal cover.

\subsec{The Strip}

To investigate the causal structure it is convenient to do a conformal
transformation to flat space. For that purpose we
introduce a coordinate $\varphi$, related to $\rho$ by
\eqn\phirho{\tan\varphi =\sinh\rho~, \qquad
-{\pi\over 2}\leq\varphi\leq{\pi\over 2}~.}
Now the metric \metglobal\ turns into
\eqn\metconf{ds^2={R^2\over\cos^2\varphi}(-d\tau^2+d\varphi^2)~,}
and by a conformal transformation we thus obtain
\eqn\conftran{ds'^2={\cos^2\varphi\over R^2}ds^2=-d\tau^2+d\varphi^2~.}
This is a two dimensional flat strip, where time runs all along the strip,
the two boundaries of $AdS_2$ are mapped to the boundaries of the strip
at $\varphi=\pm\pi/2$,
and light-like directions are 45 degrees lines in the $(\tau,\varphi)$ plain.

\subsec{Poincar\'e coordinates}

Parametrizing
\eqn\poinads{\eqalign{X_0=&{1\over 2u}\left(1+u^2(R^2-t^2)\right)~,\cr
            X_1=&{1\over 2u}\left(1-u^2(R^2+t^2)\right)~,\cr
            X_2=&Rut~,\qquad u\geq 0~,}}
one obtains the metric
\eqn\metpoin{ds^2=R^2\left(-u^2dt^2+{du^2\over u^2}\right)~.}
The coordinates \poinads\ cover one half of the hyperboloid \hyperbol\ --
the Poincar\'e patch of $AdS_2$. They are thus called the
Poincar\'e coordinates.

\subsec{Black hole coordinates}

Consider the metric
\eqn\bhads{ds^2=R^2\left(-(r^2-m)d\tau^2+{dr^2\over r^2-m}\right)~.}
We shall refer to this background as an ``$AdS_2$ black hole'' with a mass
related to $m$.
Actually, by rescaling $r$ and $t$ we see that only the sign of $m$ is
important, so w.l.g. we consider only $m=\pm 1$.
For $m=-1$, we find that the background
\eqn\bhadss{ds^2=R^2\left(-(r^2+1)d\tau^2+{dr^2\over r^2+1}\right)~,}
is equivalent to $AdS_2$ in global coordinates \metglobal\
by the coordinate transformation
\eqn\cotr{r=\sinh\rho~.}
On the other hand, for $m=1$, the background \bhads\ is
\eqn\bhadsss{ds^2=R^2\left(-(r^2-1)d\tau^2+{dr^2\over r^2-1}\right)~,}
which is identical for $r\geq 1$ to the metric in \aacc,
as can be seen by the coordinate transformation
\eqn\cotra{r=\cosh\chi~,\qquad  \tau=y~.}
Hence, for $r\geq 1$ the background \bhadsss\ covers region A in the extremal
$AdS_2$ black hole.
Similarly, $-1\leq r\leq 1$ ($r=\cos\chi$, for which \bhadsss\ is equal to \bb)
covers region B, and $r\leq -1$ ($r=-\cosh\chi$) covers region C.
The gauge field of \aacc,\bb\ in these $r,\tau$ coordinates is
\eqn\atau{A_\tau=\half(1-r)~.}
Finally, the maximal extension of this 2-d black hole covers all of $AdS_2$ --
the infinite cover of the hyperboloid.

\appendix{B}{$EBH\equiv AdS_3/\Gamma_R\equiv$ Extremal BTZ}

In this appendix we show that the CFT corresponding to the extremal 2-d
black hole is equivalent to a certain (right-moving) orbifold of
$AdS_3\equiv \tilde{SL}(2,\IR)$, studied in \refs{\ls,\strominger}.
Before KK reduction, the metric in region A is
\eqn\rega{{1\over k}ds^2=d\theta^2-\sinh^2\theta dy^2+2\sinh^2\theta dydx+
dx^2~,}
where $x$ is compact and $y,\theta$ are the same as in section 2.
After changing coordinates to
\eqn\afchco{y=\theta_L~, \quad x=\half(\theta_L +\theta_R)~, \quad
\theta={\chi\over 2}~,}
we find
\eqn\adsthree{{4\over k}ds^2=d\chi^2+d\theta_L^2+d\theta_R^2+
2\cosh\chi d\theta_L d\theta_R~,}
where $\theta_R$ is periodic (since $x$ is compact).
This background corresponds to region A of the orbifold of $SL(2,\IR)$
by
\eqn\orb{g\simeq ge^{\pi r_R\sigma_3}~,}
where $r_R$ is a constant.
This can be seen by noting that the background \rega\ is obtained from
a WZW model on $SL(2,\IR)$ with $g\in SL(2,\IR)$ parametrized as in
\param\ by
\eqn\parag{g=e^{\theta_L\sigma_3/2}g(\chi)e^{\theta_R\sigma_3/2}~,}
where $g(\chi)=e^{\chi\sigma_1/2}$ in region A, as in \abc.
Hence, the exact fully asymmetric ${SL(2,\IR)\times U(1)\over U(1)}$ quotient
CFT -- leading to the EBH (or $AdS_2$ with gauge field) upon KK reduction --
is equivalent to an orbifold of $AdS_3$.

Finally, let us note that this background is actually equivalent to the
extremal BTZ black hole~\btz. Such three dimensional black holes, with mass
$M_{BTZ}=r_+^2+r_-^2$ and angular momentum $J_{BTZ}=2r_+r_-$,
can be constructed as orbifolds of the $SL(2,\IR)$ CFT \refs{\hw,\kaloper}
by (see e.g. \hkv\ for similar notation to those we use):
\eqn\btzorb{g\simeq
e^{\pi(r_{+}-r_{-})\sigma_3}ge^{\pi(r_{+}+r_{-})\sigma_3}~.}
Hence, the orbifold \orb\ is the BTZ black hole with $M_{BTZ}=|J_{BTZ}|$.

\listrefs
\end